\pdfoutput=1
\documentclass[11pt]{article}
\usepackage[T1]{fontenc}
\usepackage[utf8]{inputenc}
\usepackage{lmodern}
\usepackage[margin=1in]{geometry}
\usepackage{microtype}
\usepackage{amsmath}
\usepackage{booktabs}
\usepackage{xcolor}
\usepackage{pgfplots}
\pgfplotsset{compat=1.18}
\usepackage{url}
\usepackage[hidelinks]{hyperref}

\newcommand{\NA}{\textemdash}

\title{How far does a hash-only sender go?\\
A pre-registered map of sender memory versus bits\\
for short text messages}
\author{Marcelo Samoilenko\\
Winserv\\
\href{mailto:msamoilenko@winserv.com.br}{\texttt{msamoilenko@winserv.com.br}}\\
ORCID: \href{https://orcid.org/0009-0008-1760-0228}{0009-0008-1760-0228}}
\date{November 2026}

\begin{document}
\maketitle

\begin{abstract}
Can a sender that only computes hashes, with a language model only at the receiver, transmit short
human text more cheaply than a classical sender trained on the same domain? We answer with one
pre-registered run on 1,000 English user utterances from Taskmaster-1. The primary endpoint is the
cost per message to deliver 100\% of the messages (hash bits, plus a 2-byte negative acknowledgement
and the raw text when the first attempt fails), paired against an in-domain PPM order-1 sender. The
hash-only sender delivered 857 of 1,000 messages on the first attempt and cost 192.8 bits per
message, 18.4 bits more than PPM order 1 (95\% bootstrap interval 10.9 to 26.3); by the
pre-registered rule it spends more and is not equivalent within one byte. As secondary results we
map measured sender memory against bits. The hash-only sender (8,190 B on a host build, keyed BLAKE2b
tag included, as in every method) was the cheapest point in bits from 8,190 B up to, but excluding,
PPM order 1 at 24,844 B, 45 to 46 bits below smaz2 and Unishox2; from PPM order 1 upwards, in-domain
senders spend fewer bits, except order 2 pruned to 16 KiB. In airtime it loses to PPM order 1 on all five links priced, to Unishox2
on four, and to raw text with a tag on two, because each failure costs an acknowledgement frame and
a fallback frame. A wider receiver search recovers half of the failures and makes the cost 7.7 bits
lower than PPM order 1, at about three minutes of receiver time per failure, and still loses in
airtime on four of five links. We report the measurement as a delimitation, with every deviation of
the run.
\end{abstract}

\section{Introduction}

Some links charge dearly per bit or per frame: LoRa, LoRaWAN at low data rates, APRS, satellite
short-burst data. We use these links only to price bits and frames; we do not claim a market for
free text on them. Language models compress text well, but a compressor that codes with a model's
predictions needs the same model at the decoder, and in practice at both ends
\cite{valmeekam2023llmzip,deletang2024}. (For LLMZip this is our inference from the scheme, not a
statement of its abstract.) We ask the opposite question: can the sender stay modest, computing only
hashes at run time, while a receiver with a language model reconstructs the message?

Slepian--Wolf coding \cite{slepian1973}, of which DISCUS \cite{pradhan2003discus} is a classical
construction, lets a decoder with side information recover a source from fewer bits than the source
alone would need. We use it only as background and inspiration. Our receiver has a prior (a language
model), not a correlated observation; the asymmetry we exploit is computational, not informational.

The question we measure is not whether the scheme ``wins''. The primary endpoint is the paired cost
to deliver every message against an in-domain PPM order-1 sender. As a secondary result we place every
sender in the plane of sender memory against bits per message to deliver every message. The analysis
was fixed before the run, and the result is reported whether or not it favours the scheme. It does
not.

Our contributions are:
\begin{itemize}
\item a lossless, tag-verified protocol in which the sender runs no model and holds no table at run
time, and a language-model receiver reconstructs the message by hash-pruned beam search, with a
negative acknowledgement and a fallback that deliver every message (Section~\ref{sec:scheme});
\item a pre-registered, single-run comparison on 1,000 human utterances against in-domain PPM
senders and two short-text compressors, with identical 40-bit overhead in every method and sender
memory measured by one instrument on one platform (Section~\ref{sec:design});
\item a negative, delimiting result: the scheme is the cheapest sender in bits only in a narrow
memory band, and it loses in airtime on most links priced (Section~\ref{sec:results}).
\end{itemize}

\section{The scheme}\label{sec:scheme}

\paragraph{Sender.} A message of $L \le 255$ bytes is cut into 2-byte chunks. For each chunk, the
sender transmits the top bits of a keyed BLAKE2b hash \cite{aumasson2013blake2}, keyed by a secret
shared by both ends, of the chunk index and of the whole prefix up to that chunk: 10 bits for chunks
that start in the first 8 bytes and 4 bits afterwards (we call this profile \texttt{10:4}). It also
sends an 8-bit length field and a 32-bit tag (keyed BLAKE2b of the whole message). For $L \ge 8$ the
message costs a deterministic, length-dependent
\begin{equation}
B_{\mathrm{scheme}}(L) = 80 + 4\left\lceil \frac{L-8}{2} \right\rceil \text{ bits},
\end{equation}
so 28 to 49 bytes cost 120 to 164 bits. The sender runs no model and holds no table at run time. It
is not free of model knowledge: the \texttt{10:4} profile was calibrated on development data with
the receiver model.

\paragraph{Receiver.} A beam search over the tokens of a language model (Qwen2.5-0.5B
\cite{qwen2024}, pinned revision, fp16 on Apple MPS), with candidates grouped by decoded byte position,
keeps only candidates whose prefix hashes match the received bits. In the primary configuration the
beam holds 64 candidates and each expansion considers the 32,768 most probable tokens. Complete
candidates are tested against the tag in score order, at most 1,024 per attempt, so a wrong message
is accepted with probability at most $1024 \cdot 2^{-32} \approx 2.4 \times 10^{-7}$ per attempt;
with the two extra attempts of secondary analysis B (Section~\ref{sec:secondary}), at most
$3 \cdot 1024 \cdot 2^{-32} \approx 7.2 \times 10^{-7}$ per message.

\paragraph{Delivering every message.} When the receiver fails, it returns a 2-byte negative
acknowledgement (NACK) and the sender falls back to the raw text ($8L + 40$ bits) or, as a secondary
variant, to Unishox2. With $d \in \{0,1\}$ indicating first-attempt delivery, the cost per message to
deliver 100\% is
\begin{equation}
C = B_{\mathrm{scheme}}(L) + (1-d)\,\bigl(16 + B_{\mathrm{fallback}}\bigr),
\qquad B_{\mathrm{fallback}} = 8L + 40 \text{ for raw text.}
\end{equation}

\section{Experimental design}\label{sec:design}

\subsection{Data and selection}

The source is Taskmaster-1 \cite{byrne2019taskmaster}, self-dialogues, user turns, released under
CC BY 4.0. The test pool is the official \texttt{dev} split (770 conversations; 768 after removing two
conversations used in development). Eligible utterances are user turns, whitespace-stripped, of 28 to
49 bytes, without line breaks or a leading \texttt{\#}, with exact duplicates removed: 2,517 of them,
and 2,376 after removing the 141 identical to some training utterance. We selected $N = 1{,}000$
(42\% of the pool) as the utterances with the smallest SHA-256 of a seeded string. $N$ was fixed for
precision (expected half-width of about $\pm 7$ bits), without any number from the pool. The
pre-registration, with the SHA-256 of the list and every frozen parameter, was committed and pushed
before any decoding; the list itself was not read before that commit. By the local repository and
run logs, the commit was made at 13:43:08, the push at 13:43:10 and the decoder process started at
13:43:47 (UTC$-3$, 5 November 2026); these are local records, not an independent timestamp.

\subsection{Baselines and fair accounting}

Every method carries the same 8-bit length and 32-bit tag (40 bits of overhead), including raw text
and the fallback. Classical senders are trained on the official Taskmaster-1 training split without
the development conversations (68,272 user utterances, about 2.4 MB): PPM-C with exclusion
\cite{cleary1984,moffat1990} of order 1, of order 2 (full, and pruned to 16, 32 and 64 KiB of tables
by a rule fixed before the data), and of order 3. The pruning rule removes order-2 context entries
in increasing order of count, with lexicographic tie-breaking, until the tables fit the target, and
recomputes the PPM-C escape from the pruned counts. We also include two general-purpose short-text
compressors, smaz2 \cite{smaz2} and Unishox2 \cite{unishox2}, and raw text. PPM bits are ideal code
lengths; a C arithmetic coder written as a measuring instrument stayed within 2 bits of them on every
development message, and we report a +2-bit termination sensitivity.

\subsection{Sender memory}

We use one definition for every sender: code plus read-only data of the complete sender, tag
included, measured as the section delta of a dead-stripped executable against an identical empty
driver (\texttt{size -m}). There is one platform: host arm64 (Mach-O), Apple clang 21, \texttt{-Os}.
This is not a microcontroller; an ARM embedded toolchain was not available. The keyed BLAKE2b
reference implementation alone is 7,720 B of code: 94\% of the hash-only sender and 31\% of PPM
order 1, but 8\% of the full PPM order 2 and under 2\% of order 3. Increments over
the tag-only sender are: hash-only scheme +366 B, smaz2 +2,536 B, Unishox2 +7,873 B, PPM order 1
+17,020 B. Static RAM is zero for the scheme and the PPM senders, 2,056 B for smaz2 and 450 B for
Unishox2.

\subsection{Endpoint and reading rule}

The primary endpoint is the mean over the $N$ messages of the paired difference between the scheme's
cost $C$ (raw fallback) and the in-domain PPM order-1 sender (ideal bits + 40). The 95\% interval is
a paired percentile bootstrap with 10,000 resamples and a fixed seed; a normal approximation is
reported as sensitivity. The interval is read by two exhaustive labels, both fixed in advance:
\begin{itemize}
\item \emph{sign}: entirely below 0, the scheme spends less; entirely above 0, it spends more;
containing 0, indeterminate;
\item \emph{equivalence}: entirely within $\pm 8$ bits (one byte, the payload granularity of the
links priced), equivalent within one byte; otherwise not.
\end{itemize}
First-attempt delivery $p$ is reported with a Wilson 95\% interval, as a component and not as a
decision. The pre-registered forecast, from 30 development utterances, was $p$ between 0.744 and
0.965 (point 0.90) and a difference between $-21.7$ and $+59.7$ bits (point $+2.3$), with the
expected reading ``sign indeterminate, not equivalent''.

\subsection{Secondary analyses}\label{sec:secondary}

Labelled as secondary and fixed in advance: the paired difference against every point of the
frontier, by the same method and labels; a Unishox2 fallback in place of raw text; airtime on five
links; \textbf{A}, an in-domain byte $n$-gram of order 5 as receiver on the same bits (beam 64, top 256
tokens, cap 1,024), to measure what the language model adds; and \textbf{B}, a wider receiver search
on the primary failures only (beam 256 and, if that fails, beam 1,024), to measure what more receiver
computation buys instead of a NACK.

\subsection{Airtime model}

Airtime is computed from per-message payload sizes, with the NACK as a 2-byte frame. LoRa time on air
follows the Semtech formula, with CRC on and explicit header assumed. Meshtastic LongFast: 250 kHz,
SF11, CR 4/5, 16-symbol preamble, 16-byte header and an assumed 4-byte protobuf envelope. LoRaWAN DR0:
13 bytes of MAC overhead per packet, 8-symbol preamble, SF10 with 11-byte application payloads
(US915) and SF12 with 51-byte payloads (EU868), with fragmentation into as many packets as needed.
APRS: AX.25 at 1200 bit/s, two digipeaters, TXDELAY 0.3 s, a message wrapper of 14 characters, binary
payloads in Base91 and raw text as characters with the tag in 5 Base91 characters. RockBLOCK: one
credit per 50 bytes or fraction. On LoRaWAN Class A a downlink only opens after an uplink, so there
the NACK costs latency rather than device airtime; we count it as a frame on every link.

\subsection{What development already showed}

All of the following used data that is now development only and entered the design, not the test.
On 30 Taskmaster-1 utterances, the scheme delivered 27 of 30 on the first attempt, with 143.1 bits per
message and 11.8 s per message on an Apple M4; the three failures were search failures, the true
prefix having fallen out of the 64-wide beam. A receiver-side hybrid (in-domain $n$-gram first, the
language model on failure) delivered 23 of 30 on 30 utterances sampled from 2,443 of the official
test split. On the 30 development utterances, the scheme's cost to deliver 100\% with a raw fallback
was 179.9 bits against 177.5 for the in-domain PPM order 1 (+2.4 bits), a difference 30 messages
cannot sign. Pruning PPM order 2 to 16 KiB of tables with the pre-specified rule produced a point
worse than PPM order 1 (264.3 bits): recomputing the PPM-C escape on a pruned context makes every novel
byte cost about 13 bits. The rule was not changed.

\paragraph{Exposure.} Before the pre-registration, an external reviewer consulted by the author computed pool-level values (PPM
order 1, 174.3 bits; scheme, 140.7 bits; break-even first-attempt delivery, 0.906). They entered no
parameter. Because $N$ is 42\% of the pool, they are close to the test values (174.4 and 140.9); we
report them so that the reader can weigh this.

\section{Results}\label{sec:results}

\subsection{Primary endpoint}

The cost to deliver 100\% with a raw fallback exceeded the in-domain PPM order 1 by
\textbf{18.4 bits per message (95\% bootstrap interval 10.9 to 26.3)}. By the pre-registered rule the
scheme \emph{spends more} and is \emph{not equivalent} within one byte. The expected reading at the
forecast point, ``sign indeterminate'', failed. The value lies inside the forecast range ($-21.7$ to
$+59.7$) but well above its point ($+2.3$). By accounting, not by a measured cause, most of the gap
follows from a first-attempt delivery of 85.7\% (857 of 1,000; Wilson 83.4\% to 87.7\%) instead of
90\%: at $p = 0.90$ the same arithmetic on the test means gives +2.5 bits. The normal approximation
gives 10.7 to 26.2; adding 2 bits of arithmetic-coder termination to PPM gives 16.4 (8.9 to 24.3). The
mean cost to deliver 100\% was 192.8 bits (bootstrap 185.0 to 201.0), of which 140.9 are the scheme's
own bits. No wrong message was accepted.

\subsection{The frontier in bits}

\begin{figure}[t]
\centering
\begin{tikzpicture}
\begin{axis}[
  width=0.86\textwidth, height=0.5\textwidth,
  xmode=log, xmin=5000, xmax=800000,
  ymin=95, ymax=365,
  xlabel={Sender memory, code + read-only data (bytes, log scale)},
  ylabel={Bits per message to deliver 100\%},
  axis lines=left, tick align=outside,
  grid=major, major grid style={line width=0.2pt, draw=black!12},
  every axis plot/.append style={line width=0.6pt},
  legend style={draw=none, fill=none, font=\footnotesize, at={(0.99,0.99)}, anchor=north east},
  legend cell align=left,
  label style={font=\small}, tick label style={font=\footnotesize},
  clip=false,
]
\addplot[const plot, draw=black!45, line width=0.6pt, mark=none]
  coordinates {(7824,344.0) (8190,192.8) (24844,174.4) (42168,143.7) (74938,140.5) (440626,114.6) (800000,114.6)};
\addlegendentry{Cheapest measured sender up to that memory}
\addplot[only marks, mark=*, mark size=2.6pt, draw=black, fill=black]
  coordinates {(8190,192.8)};
\addlegendentry{Hash-only scheme (primary)}
\addplot[only marks, mark=diamond*, mark size=3.0pt, draw=black, fill=white]
  coordinates {(8190,166.7)};
\addlegendentry{Scheme with wider receiver search (B)}
\addplot[only marks, mark=triangle, mark size=2.8pt, draw=black!60]
  coordinates {(7824,344.0) (10360,238.2) (15697,239.3)};
\addlegendentry{Raw text and short-text compressors}
\addplot[only marks, mark=square, mark size=2.2pt, draw=black!60]
  coordinates {(24844,174.4) (25786,258.6) (42168,143.7) (74938,140.5) (96462,140.5) (440626,114.6)};
\addlegendentry{In-domain PPM}
\node[font=\scriptsize, anchor=west] at (axis cs:8600,344.0) {raw text};
\node[font=\scriptsize, anchor=south west] at (axis cs:8500,194) {scheme};
\node[font=\scriptsize, anchor=west] at (axis cs:8700,166.7) {scheme + B};
\node[font=\scriptsize, anchor=south] at (axis cs:10360,242) {smaz2};
\node[font=\scriptsize, anchor=south west] at (axis cs:15697,242) {Unishox2};
\node[font=\scriptsize, anchor=north west] at (axis cs:25500,172) {o1};
\node[font=\scriptsize, anchor=west] at (axis cs:28000,258.6) {o2, 16 KiB};
\node[font=\scriptsize, anchor=south] at (axis cs:42168,147) {o2, 32 KiB};
\node[font=\scriptsize, anchor=north] at (axis cs:74938,136) {o2, 64 KiB};
\node[font=\scriptsize, anchor=south west] at (axis cs:96462,143) {o2, full};
\node[font=\scriptsize, anchor=east] at (axis cs:400000,114.6) {o3};
\end{axis}
\end{tikzpicture}
\caption{Measured sender memory (host arm64 build, keyed BLAKE2b tag included in every sender)
against mean bits per message to deliver 100\%, on the 1,000 test utterances. The grey step is the
lowest cost among the pre-registered senders that fit within a given memory; the scheme sets it from
8,190 B up to PPM order 1 at 24,844 B. The open diamond is the same sender with secondary analysis B,
which changes receiver effort only. Values are in Table~\ref{tab:frontier}.}
\label{fig:frontier}
\end{figure}
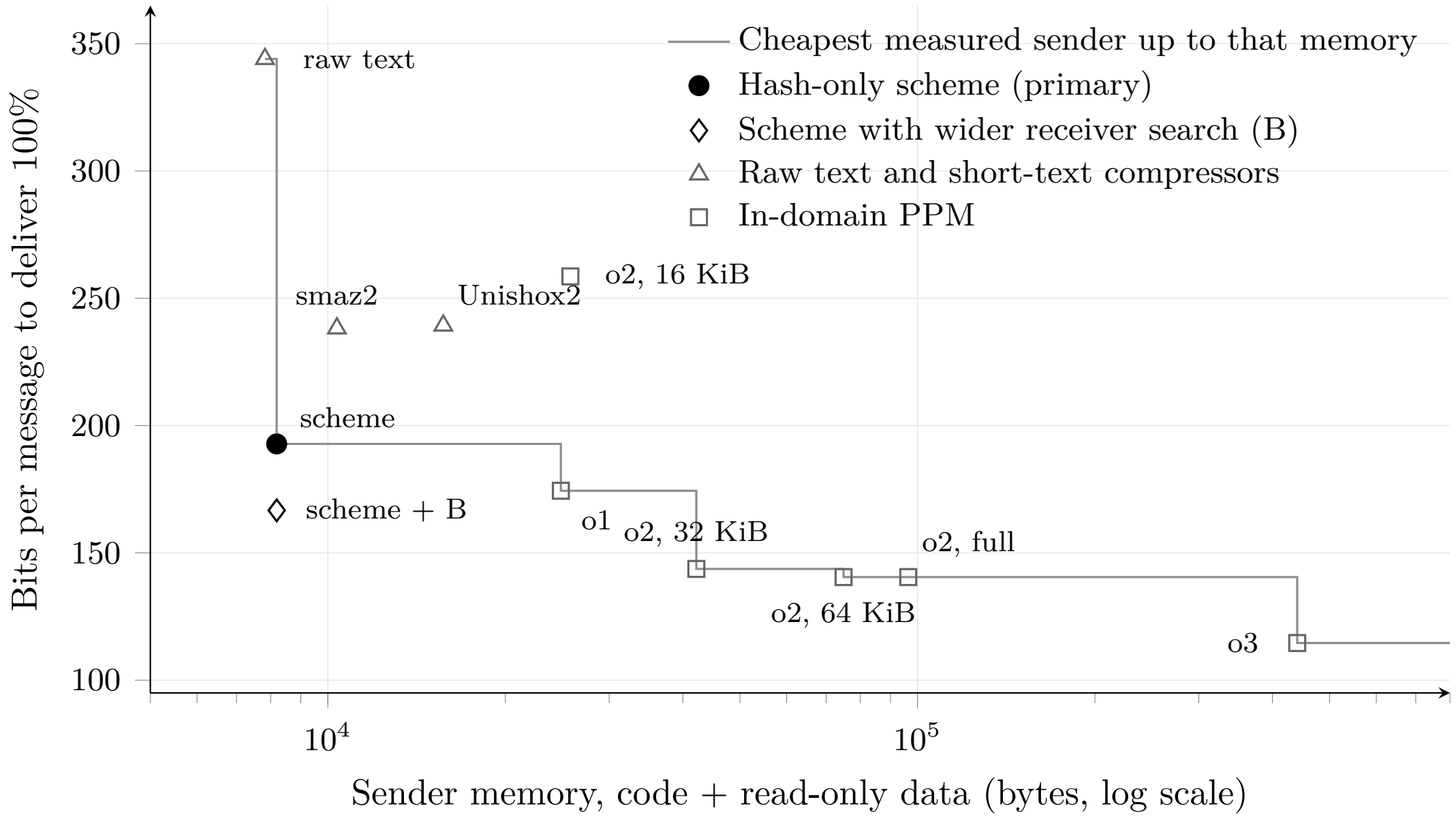

Table~\ref{tab:frontier} and Figure~\ref{fig:frontier} place every pre-registered sender in the plane
of measured memory against bits to deliver 100\%.

\begin{table}[t]
\centering
\caption{Sender memory against mean bits per message to deliver 100\% ($N = 1{,}000$; 40 bits of
overhead in every row; PPM in ideal bits). The scheme row includes the NACK and the raw fallback of
failed first attempts. Last column: paired mean of the scheme's cost minus the
point's, with 95\% bootstrap interval; negative means the scheme spends fewer bits.}
\label{tab:frontier}
\small
\begin{tabular}{lrrr}
\toprule
Sender & Memory (B) & Bits for 100\% & Scheme minus point \\
\midrule
Raw text (tag only) & 7,824 & 344.0 & $-151.1$ ($-159.1$, $-142.9$) \\
\textbf{Hash-only scheme (NACK + raw)} & \textbf{8,190} & \textbf{192.8} & \NA \\
smaz2 & 10,360 & 238.2 & $-45.3$ ($-53.1$, $-37.4$) \\
Unishox2 & 15,697 & 239.3 & $-46.4$ ($-54.2$, $-38.4$) \\
PPM order 1 & 24,844 & 174.4 & $+18.4$ ($10.9$, $26.3$) \\
PPM order 2, pruned to 16 KiB & 25,786 & 258.6 & $-65.8$ ($-74.1$, $-57.3$) \\
PPM order 2, pruned to 32 KiB & 42,168 & 143.7 & $+49.2$ ($41.8$, $56.9$) \\
PPM order 2, pruned to 64 KiB & 74,938 & 140.5 & $+52.3$ ($44.8$, $60.1$) \\
PPM order 2, full & 96,462 & 140.5 & $+52.3$ ($44.8$, $60.2$) \\
PPM order 3 & 440,626 & 114.6 & $+78.2$ ($70.7$, $86.0$) \\
\bottomrule
\end{tabular}
\end{table}

In bits, no measured point dominates the scheme from 8,190 B up to, but excluding, PPM order 1 at
24,844 B; from PPM order 1 upwards every in-domain PPM spends fewer bits, except order 2 pruned to
16 KiB, which the pre-specified pruning rule makes worse than order 1. Order 2 pruned to 64 KiB
matches the full order 2 in bits with 22\% less memory. We did not measure a PPM of order 0, which
would sit in the 8 to 25 KB band; it was not pre-registered. With a Unishox2 fallback instead of raw
text, the mean cost was 178.4 bits and the difference against PPM order 1 was +4.0 bits ($-1.4$ to
9.6): sign indeterminate, not equivalent.

\subsection{Airtime}

Table~\ref{tab:airtime} gives the mean airtime per message to deliver 100\% under the frame model of
Section~\ref{sec:design}.

\begin{table}[t]
\centering
\caption{Mean airtime per message to deliver 100\% (seconds; RockBLOCK in credits), $N = 1{,}000$. The
scheme rows include the NACK and fallback frames of failed first attempts. Order 2 is the full
model.}
\label{tab:airtime}
\small
\begin{tabular}{lrrrrr}
\toprule
Sender & Meshtastic & US915 DR0 & EU868 DR0 & APRS & RockBLOCK \\
\midrule
Scheme + raw fallback & 0.693 & 0.999 & 2.293 & 0.992 & 1.310 \\
Scheme + Unishox2 fallback & 0.679 & 0.941 & 2.214 & 0.987 & 1.286 \\
Scheme + B + raw fallback & 0.612 & 0.860 & 2.016 & 0.881 & 1.154 \\
\midrule
Raw text (tag only) & 0.719 & 1.590 & 2.700 & 0.908 & 1.155 \\
smaz2 & 0.620 & 1.156 & 2.132 & 0.868 & 1.000 \\
Unishox2 & 0.622 & 1.161 & 2.134 & 0.869 & 1.000 \\
PPM order 1 & 0.566 & 0.891 & 1.887 & 0.807 & 1.000 \\
PPM order 2 & 0.534 & 0.751 & 1.748 & 0.772 & 1.000 \\
PPM order 3 & 0.508 & 0.675 & 1.642 & 0.745 & 1.000 \\
\bottomrule
\end{tabular}
\end{table}

In airtime the scheme loses to PPM order 1 on all five links and to Unishox2 on four; it wins against
Unishox2 only on US915 DR0, whose 11-byte packets reward its deterministic size. Its advantage in bits
over Unishox2 disappears because each failure pays a NACK frame and a fallback frame. On APRS and
RockBLOCK it also loses to raw text with a tag, which needs less memory: on those links raw text
dominates the scheme in both memory and airtime.

\subsection{Secondary A: the receiver model}

With an in-domain byte $n$-gram of order 5 as receiver on the same bits, first-attempt delivery fell to
723 of 1,000 (1.55 s per message). The language model delivered 183 messages that the $n$-gram missed
and missed 49 that the $n$-gram delivered (674 delivered by both): a net gain of 134 per 1,000. With
the $n$-gram receiver, the cost to deliver 100\% exceeded PPM order 1 by 70.9 bits (60.8 to 81.2).

\subsection{Secondary B: more receiver search instead of a NACK}

Retrying the 143 failures with a beam of 256 recovered 46; retrying the remaining 97 with a beam of
1,024 recovered 26, so 929 of 1,000 were delivered without a NACK (Wilson 91.1\% to 94.3\%). The mean
cost to deliver 100\% then fell to 166.7 bits, 7.7 bits \emph{below} PPM order 1 ($-13.2$ to $-1.8$),
at 46 s per attempt at beam 256 and 191 s at beam 1,024 (slowest 402 s): about three minutes of
receiver time per failure on average, or 25.2 s per message over all 1,000. With B the scheme still
spends more than PPM order 2 pruned to 32 KiB (+23.1 bits), full order 2 (+26.3) and order 3 (+52.1),
and it still loses to PPM order 1 in airtime on four of five links; it wins only on US915 DR0 (0.860
against 0.891 s).

\subsection{Receiver cost and search diagnostics}

The primary receiver took 11.9 s per message on an Apple M4 (median 11.5 s), with up to 2.4 GB of
resident memory and 5.6 GB of GPU driver memory; secondary B reached 2.9 GB and 10.4 GB. The cap of
1,024 tag checks truncated the list of complete candidates on all 1,000 messages, and every one of the
143 primary failures exhausted the cap (142 of the 277 failures of A).

\subsection{Deviations}

The primary run was interrupted twice by an unrelated local model started on the same machine
(SIGTERM after 727 messages, SIGKILL after 729). It was resumed as the pre-registration allows: no
completed row was redone or discarded, and model revision, library versions and device were the same
before and after. Decode times of messages 650 to 728 are inflated (one took 304 s). Delivery does not
depend on load, because the decoder has no time limit and its cache budget is fixed; the failure rate
there was 11 of 79, against 132 of 921 elsewhere. A provisional analysis of the primary run and of A
was computed while B was still running; the decision to run B had been recorded before any endpoint
was computed.

\section{Related work}

We opened every source cited in this section, but for all except DISCUS (metadata only) we read
only the abstract. Language-model compressors code with the model's predictions
\cite{valmeekam2023llmzip,deletang2024}. Distributed source coding sends syndromes of a linear code
and decodes with side information \cite{pradhan2003discus}; learned decoders for distributed
compression are reviewed by Ozyilkan and Erkip \cite{ozyilkan2024}, for abstract sources and images,
not text. Merhav \cite{merhav2024} shows that a universal Slepian--Wolf rate region is attainable with
an incremental scheme and a low-rate feedback link, a theoretical counterpart of our NACK; we did not
read the original incremental scheme by Draper. The closest regime we found is a 2026 preprint that
names the ``weak-sender/strong-receiver'' setting and compresses text by deletion, with a language
model reconstructing at the receiver \cite{zou2026}; it is lossy and has no reconstruction check,
while ours is lossless and checked by a tag. Another 2026 preprint uses a language model to propose
corrections that LDPC parity then verifies \cite{park2026}; that is channel decoding, not source
coding. None of this proves novelty; our search of prior work was not systematic.

\section{Limitations}

\begin{itemize}
\item One corpus (Taskmaster-1, staged dialogues, same distribution as development), one receiver
model, one language (English), one length band (28 to 49 bytes); messages were decoded without the
dialogue context. Contamination of the receiver model by Taskmaster-1 was not measured.
\item The in-domain PPM was trained on 2.4 MB of the same corpus: the best case for the classical
sender.
\item Sender memory was measured on a host arm64 build, not a microcontroller; the BLAKE2b reference
is most of the size of the smallest senders, and a compact implementation was not measured. Non-constant tables of smaz2
and Unishox2 count as RAM in our definition; on a microcontroller their initial image would also
occupy flash.
\item The pruning rule for PPM order 2 is naive; a pruning that re-estimates escapes is future work.
No PPM of order 0 was measured.
\item Airtime depends on frame assumptions (2-byte NACK, header sizes, the NACK counted as a frame on
LoRaWAN).
\item The receiver runs in fp16 on Apple MPS; we do not claim bit-identical results on other machines.
\item There is no wire format and no cryptographic review: the BLAKE2b tag stands in for an audited
message authentication code.
\end{itemize}

\section{Conclusion}

With a NACK and a raw fallback, a sender that only computes hashes delivers every message, losslessly
and verified by a tag, while a receiver with a language model does the modelling. Measured against an
in-domain classical sender, it does not pay on our pre-registered endpoint: it spends 18.4 bits more
per message than PPM order 1. It is the cheapest sender in bits only between 8,190 B and PPM order 1 at
24,844 B (on a host build, with 7,720 B of shared BLAKE2b in both), and in airtime its failures cost frames that
erase its bit savings on most links. A wider receiver search makes it cheaper than PPM order 1 in bits,
but not in airtime, and only at minutes of receiver time per failure. We did not identify a device
class that sends typed free text and cannot hold a PPM order-1 sender; this is our inference, not a
measurement. For this regime, a small in-domain model at both ends was the better design in our
measurements. We publish the map so that claims about receiver-side language models can be weighed
against an in-domain classical sender with measured memory.

\section*{Software, data and reproducibility}

Receiver model Qwen2.5-0.5B, revision \texttt{060db6499f32faf8b98477b0a26969ef7d8b9987}. Third-party C
sources at pinned commits: BLAKE2 \texttt{ed1974ea} (CC0, OpenSSL or Apache-2.0), Unishox2
\texttt{6ec2ff07} (Apache-2.0), smaz2 \texttt{71ecc43e} (MIT). Taskmaster-1 is used under CC BY 4.0.
The pre-registration was committed as \texttt{919fd75} and the frozen code as \texttt{4013e6f}; the
per-message results, the analysis scripts and an audit script that recomputes the reported numbers
from the result files are in the public repository
\url{https://github.com/msamoilenko/hash-sender-llm-receiver}, a snapshot of the files needed to
check this article. The development repository, with the commit history cited above, the decision
log and its errata, is available from the author on request.

\section*{Use of generative AI}

This work made substantial use of language models. Claude (Anthropic), through Claude Code, was the
main assistant and orchestrated the work under the author's direction: it wrote the experiment code,
the analysis scripts and drafts of this text, and a Claude subagent reviewed the code before the
pre-registration and the manuscript before submission. Other tools served as support or reviewers:
Perplexity for literature search; Gemini, DeepSeek and ChatGPT to critique the design and the
manuscript; and a separate coding agent (Deep Code), which ran the test suite and the audit script
during development. Every suggestion was checked against the code and data and accepted or rejected
by the author. The author made the final design decisions, approved the pre-registration, and takes
responsibility for the content. No generative model is an author of this paper.

\end{document}